# Entropy-based Thermal Sensor Placement and Temperature Reconstruction based on Adaptive Compressive Sensing Theory

Kun-Chih (Jimmy) Chen, Senior Member, IEEE, Chia-Hsin Chen, Lei-Qi Wang, and Chun-Chieh Wang

***Abstract*— The high computational complexity caused by a variety of workloads in multi-core systems makes the problem of system overheating increasingly serious. Generally, a dynamic temperature management mechanism (DTM) is used to prevent system overheating. Considering the manufacturing cost and power consumption, only a small number of temperature sensors are used to obtain the temperature information on multi-core systems. The sensing temperature information is used to reconstruct the full-chip temperature distribution and leverage the DTM. Therefore, many related works have been devoted to determining the appropriate allocation of thermal sensors and temperature reconstruction methods. However, the temperature distribution of a multi-core system is often dependent on the target application, making sensor allocation challenging. Fortunately, compressive sensing (CS)-based reconstruction methods exhibit good reconstruction results, even with random thermal sensor allocation. The CS algorithm performs well while the collected signals fit Restricted Isometry Property (RIP), which is not guaranteed by using random temperature sampling. Besides, the conventional CS-based full-chip temperature reconstruction method uses fixed measurement matrix, which further worsens the efficiency of full-chip temperature reconstruction. To solve the problems, we first propose an entropy-based thermal sensor allocation to find locations that cover as many different temperature behaviors as possible. Besides, we further propose an adaptive CS-based temperature reconstruction method to dynamically adjust the involved measurement matrix, which adapts to the current temperature behavior and improves the temperature reconstruction accuracy. The experimental results show that the proposed method can reduce the average temperature reconstruction error by 18% to 95%. Besides, compared with the related works, we can improve the hardware efficiency of the proposed method by 5% to 514%.**

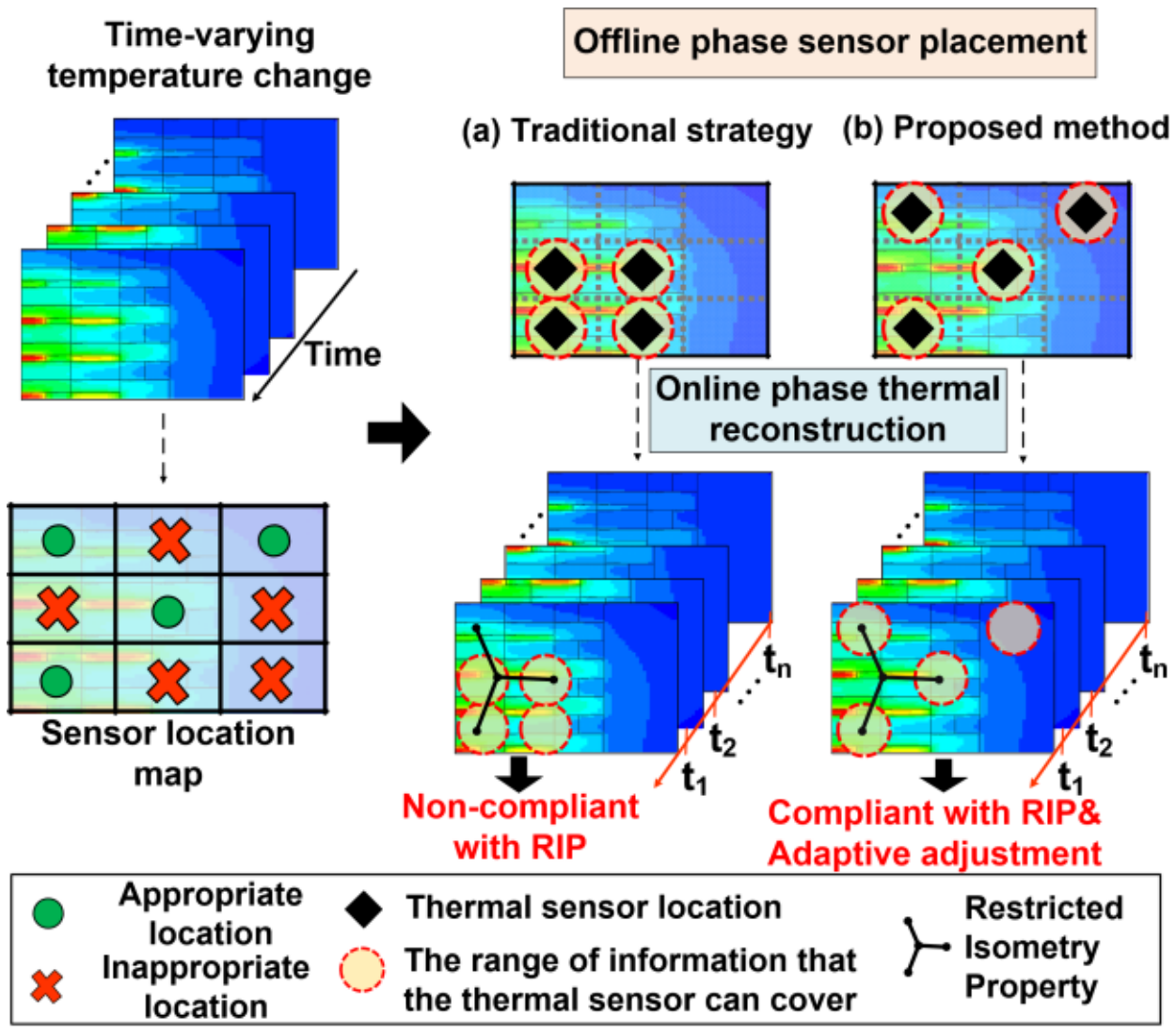

Fig. 1 The amount of temperature information covered by (a) the traditional thermal sensor allocation does not consider time-varying temperature change and RIP, whereas (b) the proposed thermal sensor allocation finds the proper locations by considering both time-varying temperature change and RIP.

## I. Introduction

Semiconductor technology has developed along with Moore's Law, which leverages the design of the multi-core system. Multi-core systems provide astonishing performance over conventional single-core systems with many computing cores. With the advancement of technology, the number of cores in multi-core systems has gradually increased to dozens or even hundreds [1]. However, the powerful processing performance of the multi-core system presents the problem of high power consumption. Owing to the increase in power consumption, the problem of thermal dissipation of chips has always been a design challenge. To prevent overheating of multi-core systems, modern heat control technologies utilize the dynamic temperature management mechanism (DTM) to control the system temperature under a certain thermal limit [2].

The efficiency of the adopted DTM is characterized by the accuracy of the information about the temperature distribution on the system. Allocating a massive number of thermal sensors on the system is the simplest way to obtain precise information on temperature distribution. However, it incurs high manufacturing costs and large area overhead [3]. When considering the manufacturing cost, the number of thermal sensors involved is usually limited in practical systems.

This work was supported by the Ministry of Science and Technology, TAIWAN, under Grant NSTC 113-2640-E-A49-005 and 114-2221-E-A49-170-MY3

K.-C. Chen and C.-C Wang are with the Institute of Electronics, National Yang Ming Chiao Tung University, Hsinchu, Taiwan (e-mail: kcchen@nycu.edu.tw; ccwang@ceres.iee.nycu.edu.tw).

C.-H. Chen and L.-Q. Wang are with the Department of Computer Science and Engineering, National Sun Yat-Sen University, Kaohsiung, Taiwan. E-mail: {dolly0104, laichi609}@cereal.cse.nsysu.edu.tw. (Corresponding Author: Kun-Chih (Jimmy) Chen)

Unfortunately, the sensor allocation problem has been proven to be an NP-hard problem [5]. The improper locations for thermal sensor allocation may lead to large estimation error on full-chip temperature reconstruction [6], as shown in Fig. 1(a). To address the problem, Reda et al. allocated thermal sensors on the hotspot area after tracking the temperature distribution under different workloads [4]. In [2], Nowroz et al. employed the minimum cutting algorithm to recursively partition the target multi-core platform into several regions and allocate thermal sensors on the energy center point of each cutting area. Although the aforementioned methods reduce the search complexity to determine the sensor locations, they still suffer from large full-chip temperature estimation errors because the temperature distribution of multi-core systems is time-varying [7]. Chen et al. employed a statistical method to analyze the correlation of the temperature change along with time to consider the time-varying temperature change when allocating thermal sensors [7]. Subsequently, the authors allocated thermal sensors to the locations where the temperature change was highly correlated with other locations. However, their method does not consider the temperature variation caused by diverse applications. Therefore, the authors further applied compressive sensing (CS) theory to randomly allocate the thermal sensors and reconstruct the full-chip temperature distribution by using several CS-based reconstruction methods [8]. The advantage of the CS-based method is that it can be applied to general-purpose multi-core systems (i.e., application independence). The requirement to make the CS algorithm perform well is that the collected signals fit the Restricted Isometry Property (RIP). However, the random sensor allocation strategy cannot guarantee the RIP. Therefore, the traditional CS-based method still suffers from large full-chip temperature reconstruction errors.

In addition to the thermal sensor allocation, it is also critical to estimate the full-chip temperature distribution by using the number-limited temperature sensing data at runtime. Because the temperature status of each location on the chip is correlated with the temperature status of other locations. Hence, the linear combinational model is usually adopted to reconstruct the full-chip temperature distribution [2][9][10][11]. However, when the target application at runtime is not based on prior data, these methods suffer from a large error in the temperature estimation of the entire chip. On the other hand, the CS theory can effectively reconstruct system information using a small amount of data [12]. Therefore, the temperature reconstruction based on CS theory can significantly reduce the influence of prior data. Chen et al. applied the CS theory to assign thermal sensors and dynamically estimate the temperature distribution [13]. However, highly complex reconstruction algorithms lead to long computing latency to reconstruct the full-chip temperature distribution, which makes it difficult to use them in real-time multi-core systems. Among previous CS theoretical reconstruction methods, computational reduction methods based on greedy algorithms have effectively reduced computational complexity, such as Orthogonal Matching Pursuit (OMP) and Stage-wise Orthogonal Matching Pursuit (StOMP) [14][15]. However, OMP and StOMP suffer from long computing latency when the number of sensor locations to be analyzed is significant. To solve the above problems, Chen et al. proposed MIB-OMP and MIB-StOMP full-chip temperature reconstruction [16]. It adopts the matrix inversion bypass feature to replace the least square method with the most complex operation. Therefore, the reconstruction error of the above-mentioned compressive sensing reconstruction method is much smaller than that of the traditional CS-based reconstruction method. As previously mentioned, the random sensor placement strategy used does not guarantee the RIP. Consequently, if the sensor locations are not appropriate after random thermal sensor placement, the full-chip temperature reconstruction error remains significant.

To solve the aforementioned problems, we employ the entropy information to analyze the temperature distribution on the chip. The entropy information is assessed after each iteration to guide the selection of locations for thermal sensor placement. Higher entropy indicates that the allocated thermal sensors effectively capture the diverse features of temperature behavior in the multi-core system, and vice versa. Therefore, we can find a sensor allocation way with the highest entropy to close the RIP. Since each location search result is a subset of a random search, we can employ CS-based reconstruction methods from [16] to reconstruct the full-chip temperature distribution. However, the CS-based reconstruction method is based on the pre-defined measurement matrix, which cannot adapt to time-varying temperature behavior. Therefore, we propose an adaptive measurement matrix adjustment method based on the temperature estimation error at each temperature reconstruction iteration, as shown in Fig. 1(b). The contributions in this work are summarized below.

1) We employ information theory to propose an entropy-based thermal sensor placement method for multi-core system platforms.
2) We integrate the use of an adaptive filter to propose an LMS-based adaptive CS-based temperature reconstruction method.
3) We develop a hardware architecture design for the LMS-based adaptive compressive sensing method to meet real-time temperature reconstruction needs, which is advantageous for thermal-aware multi-core systems.

We combine the Hotspots thermal simulation tool [17] with the Sniper x86-64 multi-core simulator [18] to simulate the thermal behavior of the multi-core processor architecture and evaluate the proposed methods. Moreover, we evaluate our proposed methods on a practical FPGA with a multi-core platform. Experimental results indicate that the proposed entropy-based thermal sensor allocation method and the adaptive CS-based temperature reconstruction method can reduce the average temperature reconstruction error by 18% to 95% compared with related works. Finally, the proposed method can also improve the hardware efficiency by 5% to 514% compared with the conventional non-adaptive temperature reconstruction method.

The rest of the paper is described as follows. In Section II, we explore state-of-the-art methods, including thermal sensor allocation techniques and full-chip temperature reconstruction. Section III introduces our proposed entropy-based thermal sensor allocation method, along with an adaptive compressive (CS)-based approach for full-chip temperature reconstruction, and outlines the corresponding hardware architecture. In Section IV, we present and discuss the experimental results to evaluate the effectiveness of our proposed methods. Finally, we conclude our work in Section V.

## II. Background and Related Works

### A. Thermal sensor placement based on energy-greedy [2]

To determine suitable locations to place a limited number of thermal sensors, Nowroz et al. proposed a minimum cutting algorithm to recursively divide several regions and place thermal sensors on the energy center point of each cutting region. Accordingly, two sensor placement methods are based on the minimum cutting algorithm: 1) energy-center sensor allocation and 2) energy-cluster sensor allocation. The energy-center sensor allocation method entails placing the thermal sensor in the geometric center of the cutting area, aiming to cover all the temperature information in the area. However, it is rare for the hotspot to be in the center of the cutting area, which causes significant reconstruction errors. The energy-cluster sensor allocation method utilizes the K-means clustering algorithm to place potential hotspots. Subsequently, when the energy of the cutting region is greater than the threshold, the next cutting will be performed. However, this placement method emphasizes hotspot consideration, which tends to overestimate the overall system temperature.

### B. An interpolation-based thermal sensor placement [19]

Since hotspots on multi-core systems are the key to activating dynamic temperature management mechanisms, the information on temperature reconstruction is usually used to manage the hotspot temperatures. In [19], Long et al. proposed a method that uses a large number of sensor placements to locate hotspots accurately. The first step involves having all the sensors evaluate the approximate location of the hotspots. Subsequently, high-precision sensors are allowed to predict the location and temperature of the hotspots further. An interpolation-based approach is then proposed to estimate information related to the on-chip temperature distribution. However, the temperature distribution of multi-core systems is time-varying at runtime. This causes the hotspot locations to differ at each moment, resulting in estimation inaccuracies. Additionally, the interpolation method assumes a linear temperature gradient between thermal sensors. The temperature gradient around the hottest points often changes significantly, which further increases the estimation error.

### C. A principal component analysis-based thermal sensor placement [9]

To determine the best thermal sensor placement, Ranieri et al. proposed randomly selecting the thermal simulation map for principal component analysis (PCA) to generate a linear model of the system's thermal behavior. Furthermore, the authors utilize a greedy algorithm to select thermal sensor locations based on the linear model. However, the linear models generated by PCA are highly dependent on the prior thermal map. Hence, the approach requires an extensive and comprehensive prior thermal map. In addition, the efficiency of temperature estimation is highly dependent on the linear models involved. Consequently, PCA-based methods suffer from large temperature estimation errors.

### D. CS-based thermal sensor placement for thermal-aware NoC system [13]

Compressive sensing (CS) can effectively reconstruct system information using a small amount of data. Owing to the sparse temperature characteristic in the frequency domain, Chen et al. apply CS theory to allocate thermal sensors and estimate the complete on-chip temperature distribution. Subsequently, the real-time thermal map of the system is dynamically reconstructed using the Stagewise Orthogonal Matching Pursuit (StOMP) method. However, the computing complexity of StOMP is increased with respect to the increasing number of positions where the thermal sensor can be placed. Therefore, it is difficult to reconstruct the full-chip temperature efficiently at runtime.

### E. A low-complex CS-based thermal reconstruction for multi-core systems [16]

As the complexity of multi-core systems grows, the large workload diversity leads to serious thermal issues, and the number of thermal sensors placed is usually limited due to manufacturing costs. As mentioned before, CS theory has proven efficient in reconstructing original signals using fewer sampling data, but its high computational complexity makes it unsuitable for real-time temperature monitoring in current multi-core systems. Chen et al. first proposed a grid-based sensor placement approach to place thermal sensors on the target multi-core system. Besides, the authors adopt the Matrix Inversion Bypass (MIB) property to reduce the computational complexity of the matrix inversion to reconstruct the original signals. However, the random thermal sensor allocation strategy involved cannot guarantee the RIP. Besides, the fixed measurement matrix involved is not suitable for the time-varying temperature behavior on multi-core systems at runtime. Therefore, the efficiency of the full-chip temperature reconstruction still needs to be improved.

## III. Proposed Entropy-based Thermal Sensor Placement and Temperature Reconstruction

### A. Entropy-based thermal sensor placement

As mentioned before, the CS theory can assist with the efficient system thermal map reconstruction by using a small number of thermal sensors [13][16]. However, since the conventional CS-based method uses a random sensor placement method, we cannot guarantee that the collected sensing data fits the RIP, which still results in a large full-chip temperature distribution reconstruction error. The RIP in CS ensures that the collected information has minimal redundancy and is distinguished. On the other hand, the entropy in information theory is used to quantify the uncertainty or randomness of the collected data. A high-entropy signal corresponds to a distribution where outcomes are nearly equiprobable, reflecting minimal redundancy and maximal information content. Motivated by this principle, we incorporate the concept of entropy, defined in information theory, into the thermal sensor placement problem in this work.

The entropy is a quantitative criterion that is commonly used

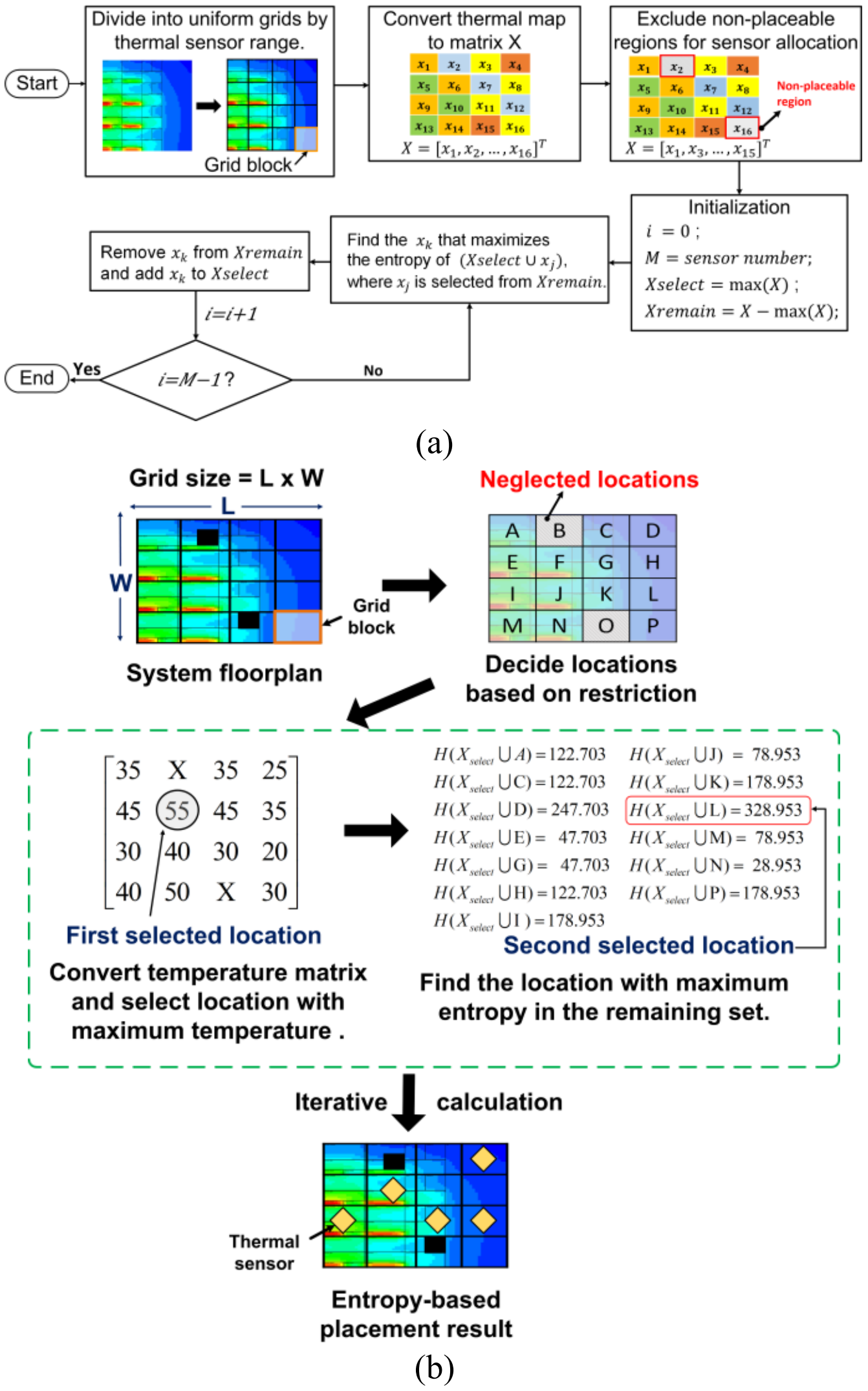

Fig. 2 (a) The flowchart of the entropy-based thermal sensor placement process, and (b) the example to find proper thermal placement locations based on the proposed approach.

in information theory to judge the amount of information in a system. The higher entropy means that we can consider more information from the data collected. In other words, the entropy is used to represent the chaotic degree of information. Among the different entropy forms, Shannon entropy is the most widely used entropy measure in information theory [29]. Shannon entropy assumes no specific underlying distribution and is effective for a wide range of scenarios. Because we need to capture the overall uncertainty and diversity of temperature behaviors across the chip, Shannon entropy can provide a well-established and interpretable metric for the sensor allocation. Therefore, we employ Shannon entropy in this work, which can be defined as

$$H(X) = \Sigma_{x \in X}\, p(x) \log_2\left(\frac{1}{p(x)}\right), \quad (1)$$

where $p(x)$ represents the probability of element $x$, and $H(X)$ represents the entropy value of the entire set of element $x$.

To facilitate the entropy analysis, similar to [13], we adopt a grid-based strategy to place thermal sensors. By providing the target multi-core system platform and the constraints of a predefined $K$-by-$K$ sized grid, $K^2$ grid blocks are considered to place thermal sensors. Because some regions (e.g., memory cells) on a multi-core system are too critical to place any thermal sensors, the involved grid-based strategy can exclude these grid blocks, covering these restricted regions, to place thermal sensors. Fig. 2 shows the flowchart of the proposed entropy-based thermal sensor placement method. In this work, the temperature distribution on the multi-core system means the data distribution in $X$, which can be represented as

$$X = [x_1, x_2, \ldots, x_{K^2}]^T, \quad (2)$$

where $x_i$ denotes the temperature at the $i$-$th$ grid block and $K^2$ represents the total number of grid blocks of interest. It is evident from (1) that we need a probability density function (PDF) to aid the calculation of entropy. From the analysis of a large number of thermal maps, the data distribution in $X$ (i.e., (2)) is a multivariate normal distribution. Therefore, we use a multivariate normal distribution probabilistic model. The probability density model of the multivariate normal distribution is defined as

$$pdf(X) = \frac{1}{\sqrt{(2\pi)^N \det(\sum X)}} \times \exp[-\frac{1}{2}(X - \mu_X)^T (\sum X)^{-1} (X - \mu_X)], \quad (3)$$

where $pdf(\cdot)$ represents the PDF of a random variable; $det(\cdot)$ represents the determinant of a matrix; $\mu_X$ and $\Sigma X$ denote the mean value; $N$ represents the number of grid blocks we involve (i.e., the $K^2$ in (2)), and the covariance matrix of the multivariate random variable set $X$, respectively. The probability function in (3) is used to replace the $p(x)$ in (1) and obtain

$$H(X) = -\int pdf(X) \times \log(pdf(X)) \times dX, \quad (4)$$

where $H(X)$ denotes the entropy of matrix $X$. By using the Mahalanobis Transformation Lemma [20], the formula (4) can be further simplified to

$$H(X) = \frac{N}{2} \times (1 + \log(2\pi)) + \frac{1}{2} \times \log(\det(\Sigma X)). \quad (5)$$

The entropy $H(X)$ in (5) is different in terms of the number of elements in each set and the covariates. The covariate between sets indicates the temperature difference between each location of interest on the chip. If each thermal sensor can monitor as much temperature information as possible, it can cover more temperature behavior information in the system. In other words, more information can be acquired on the temperature of the system using fewer thermal sensors.

According to (5), we can obtain information regarding the thermal entropy according to the current thermal sensor allocation. Furthermore, our goal is to find thermal sensor allocations that cover as much temperature behavior as possible on a multi-core system. Therefore, it is necessary to find a set of thermal sensor allocations to maximize thermal entropy, which can be found by using the flowchart in Fig. 2(a). To apply entropy analysis, we divide the thermal map into several grid blocks, resulting in a temperature matrix, $X$, after the thermal simulation. Thermal sensors should be placed in grid blocks, which help to maximize entropy. Therefore, we begin by positioning the first sensor in the block with the highest temperature, as this contributes the most to entropy. We then iteratively select the grid block temperature that yields the highest entropy at each placement stage, recalculating the new entropy for the remaining blocks by using (5). After $M$-1 iterations of these calculations, we identify $M$ grid blocks for the placement of $M$ thermal sensors. Additionally, if multiple grid blocks yield the same entropy level, the selection among them will be made randomly.

In the example shown in Fig. 2(b), we consider the presence of non-placeable regions on the platform, such as memory blocks. Our goal is to place five thermal sensors on a 4-by-4 grid. First, we divide the thermal map into 16 grid blocks and identify the locations for the first thermal sensor. Since block F has the highest temperature, we will place the first thermal sensor in this grid block. Next, we will use the temperature from block F to calculate the corresponding entropy based on the temperatures of the other grid blocks, which will help us determine the second sensor placement (i.e., block L in this case). We will continue this process to find placements for the remaining thermal sensors. This method ensures that we effectively cover a wide range of temperature behaviors, even with a limited number of thermal sensors.

### *B. Proposed LMS-based Adaptive CS-based Full-chip Temperature Reconstruction Algorithm*

After allocating thermal sensors to the determined locations using the proposed entropy-based thermal sensor placement, it is necessary to reconstruct the full-chip temperature distribution precisely. As mentioned before, the MIB-CS-based method [16] brings efficient performance in reconstructing the full-chip temperature distribution to reduce the computational complexity of the traditional OMP and StOMP algorithms. However, this method involves a fixed measurement matrix to reconstruct the full-chip temperature distribution. As mentioned before, the temperature distribution of multi-core systems is usually time-varying. Consequently, it is not proper to use a fixed measurement matrix to reconstruct the full-chip temperature distribution at runtime [21] because it may lead to large full-chip temperature distribution estimation errors. Obviously, the measurement matrix should be adjusted according to the current signal (i.e., temperature information) to address the problem of the improper measurement matrix.

To solve the aforementioned problem, we consider using the LMS adaptive filter theory [22] to dynamically update the measurement matrix. Fig. 3 shows the proposed LMS-based adaptive CS-based thermal reconstruction mechanism. Initially, the measurement matrix $\Phi_{M\times N}$ is a Gaussian random matrix [13]. The transformation matrix $\Psi_{N\times N}$ is used to transform the original non-spare signals into the spare signal, which is one of the criteria for employing the CS method. Note that the $\Psi_{N\times N}$ is fixed at runtime. On the other hand, $M$ means the number of thermal sensors we will place, and $N$ represents the number of grid blocks we involve (i.e., the $K^2$ in (2)). First, we use the temperature information collected by thermal sensors and

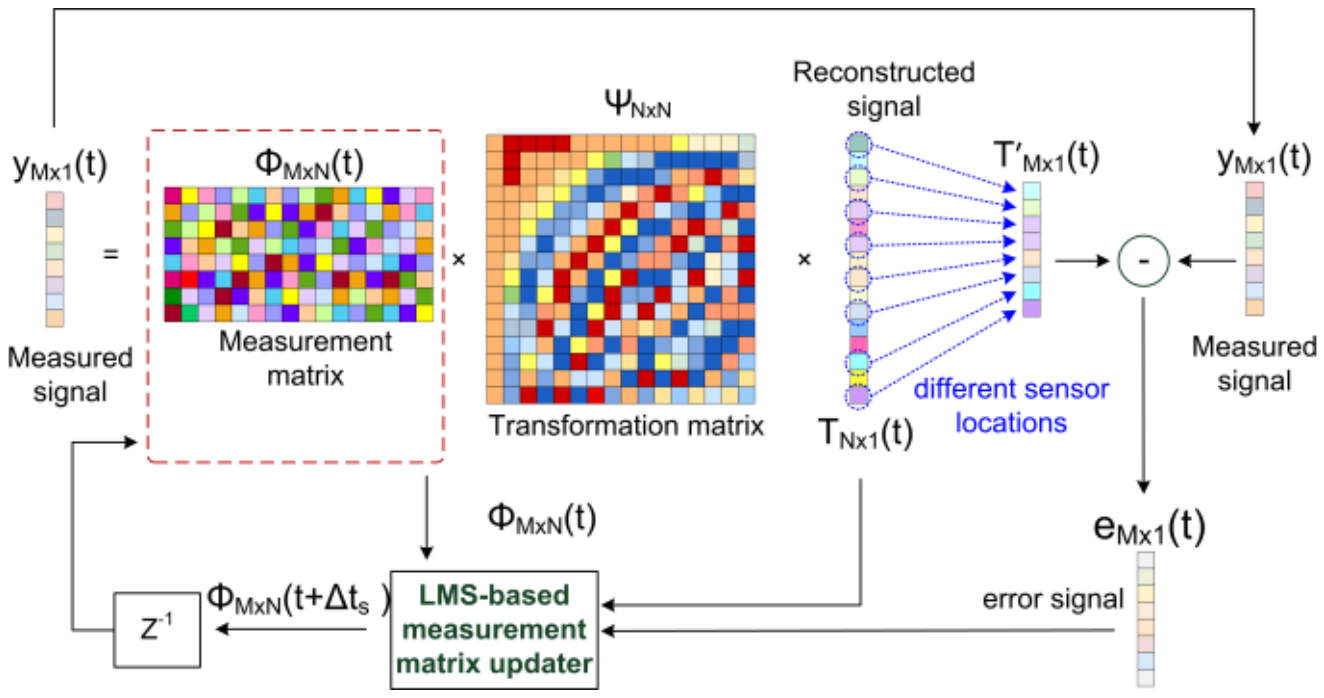

Fig. 3 Proposed LMS-based adaptive CS-based thermal reconstruction mechanism.

**Algorithm:** LMS-based Adaptive CS-based Temperature Reconstruction Algorithm

**Input:** Measured signal $y(t)\in R^{M\times 1}$, Measurement matrix $\Phi(t)\in R^{M\times N}$

Step size $\mu\in R$, Information about sensor locations $SL\in N^{M\times 1}$,

Reconstructed signal at sensor locations $T'(t)\in R^{M\times 1}$

**Output:** Updated measurement matrix $\Phi(t+\Delta t_s)\in R^{M\times N}$,

Reconstructed signal $T(t)\in R^{N\times 1}$

**Initialize:** Reconstructed signal $T$(t) = 0, Error signal $e$(t) = 0, // $e(t)\in R^{M\times 1}$

Reconstructed signal at sensor locations $T'(t)=0$

1: $T$(t) = MIB-CS-based thermal reconstruction ($y$(t), $\Phi$(t)) // the work in [16]

2: $T'(t)=T(t)[SL]$ //extract thermal signal at sensor location

3: $e(t)=T'(t)-y(t)$ //calculate the reconstruction error

4: $\Phi(t+\Delta t_s)=\Phi(t)+\mu\cdot e(t)\cdot T^T(t)$ //update the involved sensing matrix

Fig. 4 The pseudo-code of the proposed LMS-based adaptive CS-based thermal reconstruction algorithm.

$\Phi_{M\times N}$ to perform the MIB-CS-based temperature reconstruction in [16]. The MIB-CS-based temperature reconstruction method iteratively updates the inverse matrix instead of performing direct matrix inversion in the traditional CS-based temperature reconstruction methods (i.e., OMP or StOMP methods). Then, we can obtain the reconstructed temperatures at various locations on the target multi-core system. Afterward, we will compare the information about the actual measured temperature and the reconstructed temperature on the locations with thermal sensors to compute the reconstruction errors $e_{M\times 1}(t)$. Then, the LMS-based measurement matrix updater will adjust the $\Phi_{M\times N}$ based on the information about the currently involved measurement matrix and the reconstruction errors on the locations with thermal sensors. Fig. 4 shows the pseudo-code of the proposed adaptive CS-based full-chip temperature reconstruction method. After performing the MIB-CS-based temperature reconstruction in [16], the reconstructed temperature matrix at time $t$ (i.e., $T_{N\times 1}(t)$) can be obtained. Then, we will select $M$ elements from $T_{N\times 1}(t)$ to extract the temperatures at the locations where $M$ thermal sensors are placed (i.e., $T'_{M\times 1}(t)$). Therefore, we can calculate the temperature reconstruction errors at these locations (i.e., $e_{M\times 1}(t)$). By using the $e_{M\times 1}(t)$, the involved measurement matrix at the time $t+\Delta t_s$ (i.e.,

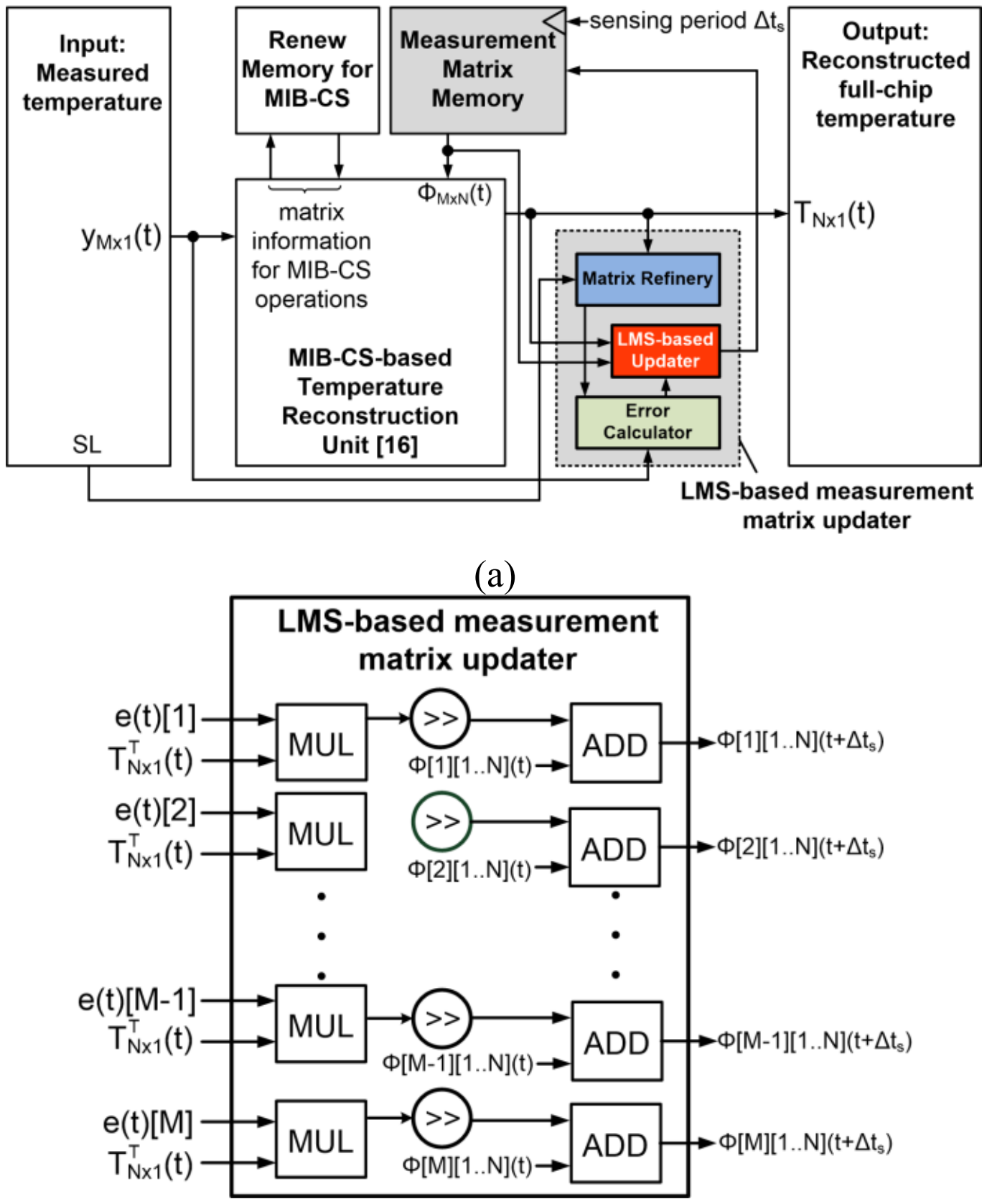

(b)

Fig. 5 (a)The architecture of the proposed LMS-based adaptive CS-based full-chip thermal reconstruction unit; (b) the proposed LMS-based matrix updater.

$\Phi_{M\times N}(t+\Delta t_s)$ can be adjusted by using the LMS adaptive filter theory, which can be formulated to

$$\Phi_{M\times N}(t+\Delta t_s)=\Phi_{M\times N}(t)+\mu\cdot e_{M\times 1}(t)\cdot T^{T}_{N\times 1}(t), \qquad (6)$$

where the $\Delta t_s$ is the sensing period of the involved thermal sensors. Note that the hyperparameter $\mu$ is the step size parameter that controls the update rate of the measurement matrix adjustment. The improper $\mu$ may affect the convergence speed. To address this problem, the precision of the $\mu$ should align with the precision of the involved thermal sensor's resolution. In this way, we can make the involved MIB-CS-based method adapt to environmental changes, thereby improving the temperature reconstruction efficiency significantly.

### *C. The architecture of the proposed adaptive CS-based temperature reconstruction unit*

To realize the proposed adaptive CS-based temperature reconstruction unit, we refer to the MIB-based temperature reconstruction method in [16]. Fig. 5(a) shows the architecture of the proposed LMS-based adaptive CS-based full-chip thermal reconstruction unit. The white blocks are the traditional MIB-CS-based temperature reconstruction unit, which can perform either the MIB-OMP method or the MIB-StOMP method. Because the MIB-CS-based temperature reconstruction method employs the recursive matrix inversion to reduce the computational complexity of the OMP and StOMP methods, the *Renew Memory* block is used to store the information on the currently involved measurement matrix for the MIB operations. On the other hand, the gray blocks are used to update the measurement matrix by following the LMS-based adaptive filter theory, as shown in the algorithm in Fig. 4.

The inserted LMS-based measurement matrix updater comprises a *Matrix Refinery*, an *Error Calculator*, and one *LMS-based Updater*. The *Matrix Refinery* extracts the information about the reconstructed temperature at the sensor locations from the reconstructed temperature $T_{N\times 1}(t)$ and obtains the $T'_{M\times 1}(t)$ in the second step of Fig. 4. Afterward, the *Error Calculator* will calculate the reconstruction error (i.e., $e_{M\times 1}(t)$ in Fig. 4) by receiving the $T'_{M\times 1}(t)$ and the currently measured temperature $y_{M\times 1}(t)$ . At last, the LMS-based updater will adjust the currently involved measurement matrix $\Phi_{M\times N}(t)$ to become $\Phi_{M\times N}(t+\Delta t_s)$ by using (6), which will be stored in the *Measurement Matrix Memory*. Note that the updated measurement matrix will be used to calculate the reconstructed temperature at the next temperature sensing time. Fig. 5(b) shows the architecture of the LMS-based updater. As mentioned before, the precision of the $\mu$ should align with the precision of the involved thermal sensor's resolution. In this work, we refer to the thermal sensor design in [28], and the resolution of this target thermal sensor is 0.32°C. Hence, we can use a shift to move the 10 bits to the right to approximate $\mu$ and consider the precision of the target thermal sensor's resolution. Therefore, the second term (i.e., $\mu\cdot e_{M\times 1}(t)\cdot T^{T}_{N\times 1}(t)$) in (6) can be calculated by shifting each element of the product vector of the currently reconstructed temperature $T^{T}_{N\times 1}(t)$ in (6) and the reconstruction error $e_{M\times 1}(t)$ in (6) right by 10 bits. Finally, the updated measurement matrix can be obtained by adding the shift-right result and the currently involved measurement matrix.

## IV. Experimental Results and Discussion

To validate the proposed entropy-based thermal sensor placement and adaptive CS-based temperature reconstruction methods, we utilize a combination of the Sniper x86-64 multi-core simulator [18] and the HotSpot tool [17]. The Sniper simulator employs the Nehalem 4-core microarchitecture floorplan and integrates the McPat-1.0 [23] power modeling framework, facilitating the calculation of power consumption for individual microprocessor components. By tracking the accumulated power consumption over a specified time interval (i.e., 1 millisecond in this study), the HotSpot tool can derive corresponding temperature data, reflecting the real-time transient temperatures of each microprocessor component under consideration. Additionally, we utilize the SPLASH-2 benchmark suite [25] in the following experiments.

### *A. Performance Comparison for Different Sensor Allocation*

To evaluate the effect of our proposed sensor placement method, we will use the average and maximum reconstruction

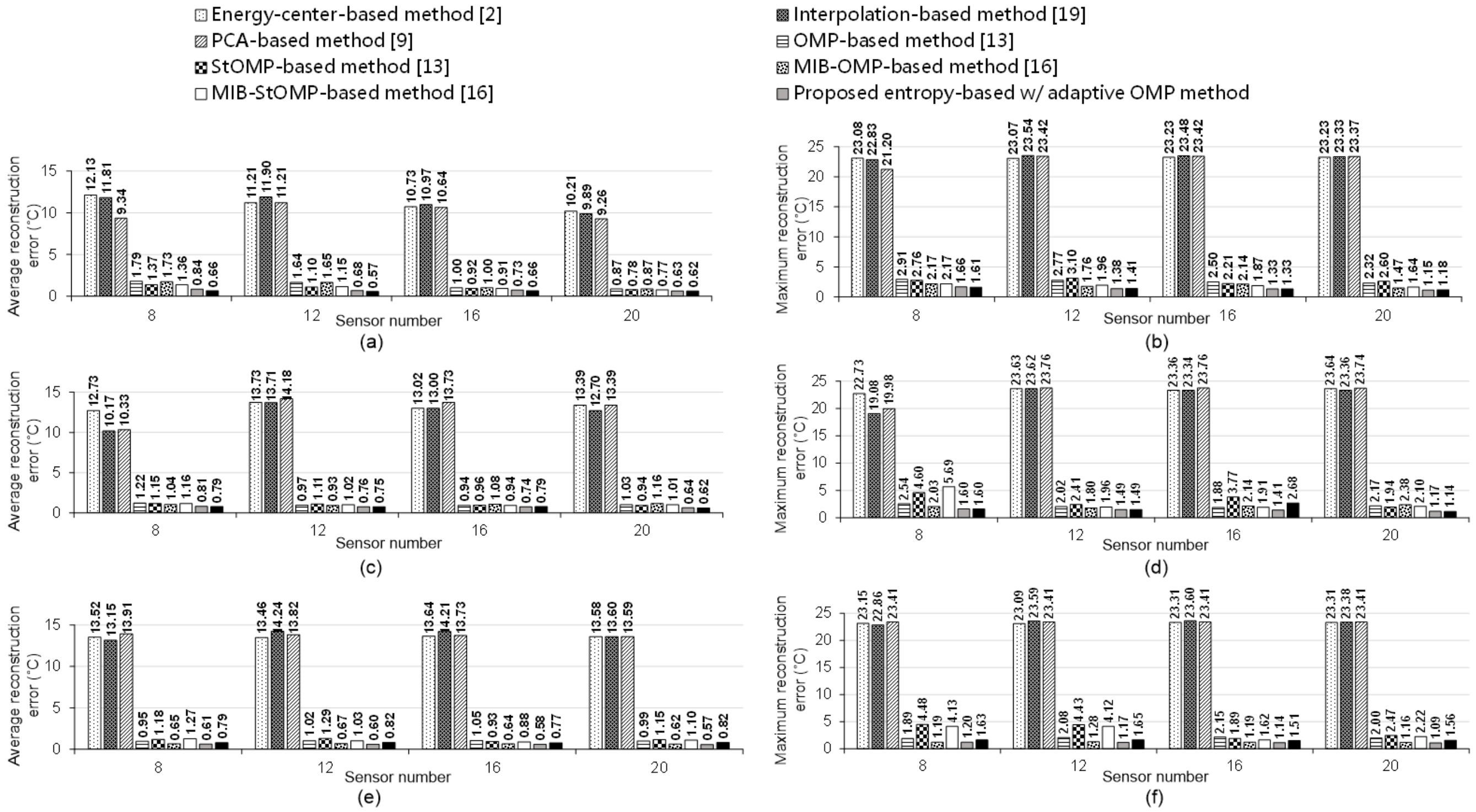

Fig. 6 The average temperature estimation error under an (a) 8×8 grid, (c) 16×16 grid, and (e) 32×32 grid; the maximum temperature estimation error under an (b) 8×8 grid, (d) 16×16 grid, and (f) 32×32 grid in a four-core after executing the entire SPLASH-2 benchmark suite.

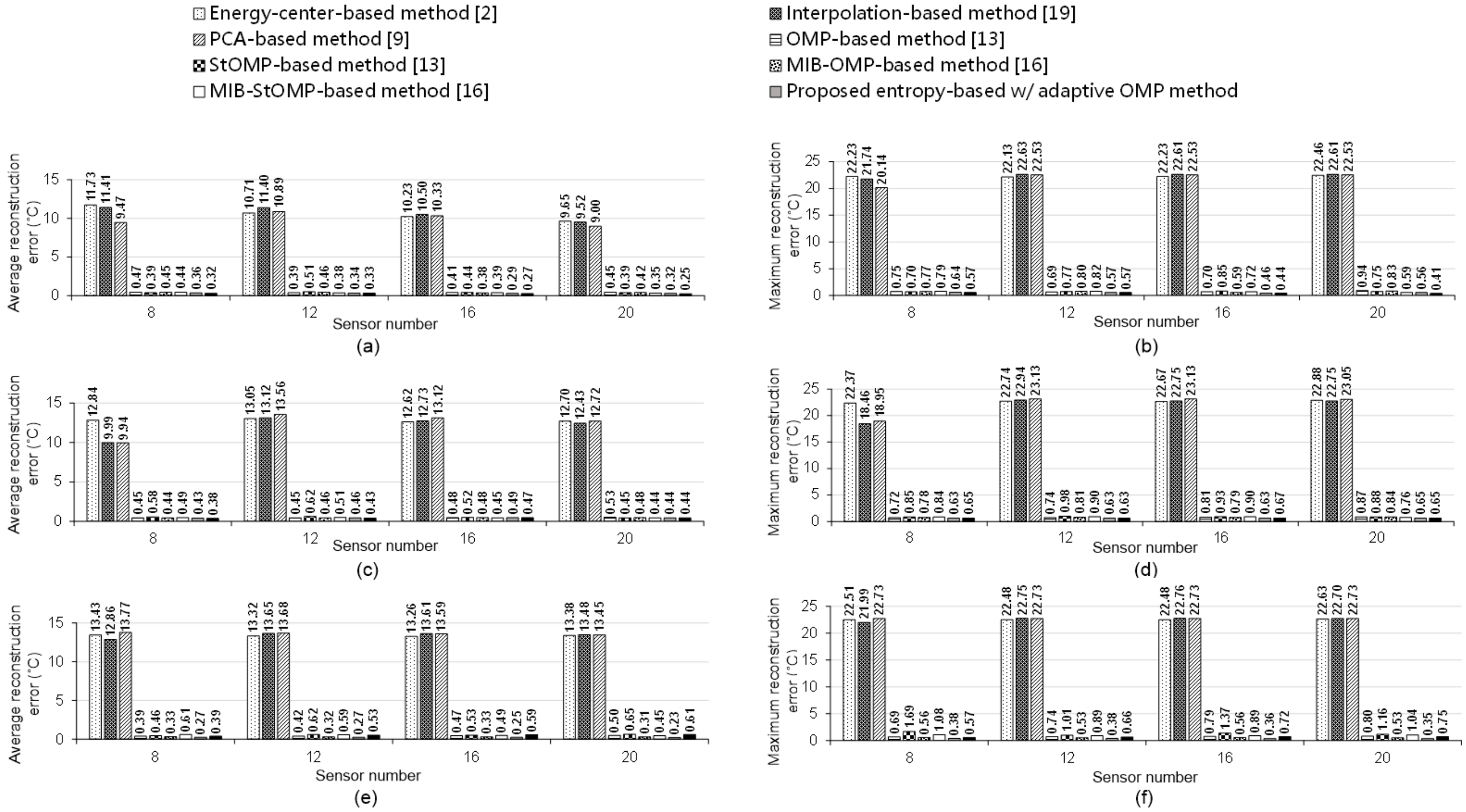

Fig. 7 The (a) average temperature estimation error under an (a) 8×8 grid, (c) 16×16 grid, and (e) 32×32 grid; the maximum temperature estimation error under an (b) 8×8 grid, (d) 16×16 grid, and (f) 32×32 grid in a four-core after executing the entire PARSEC benchmark suite.

errors under the *SPLASH-2* benchmark when 8, 12, 16, and 20 thermal sensors are placed. On the other hand, to evaluate the proposed entropy-based sensor allocation performance, we implement six different methods: 1) energy center-based method [2], 2) interpolation-based method [19], 3) PCA-based method [9], 4) CS-based method [13], 5) MIB-CS-based

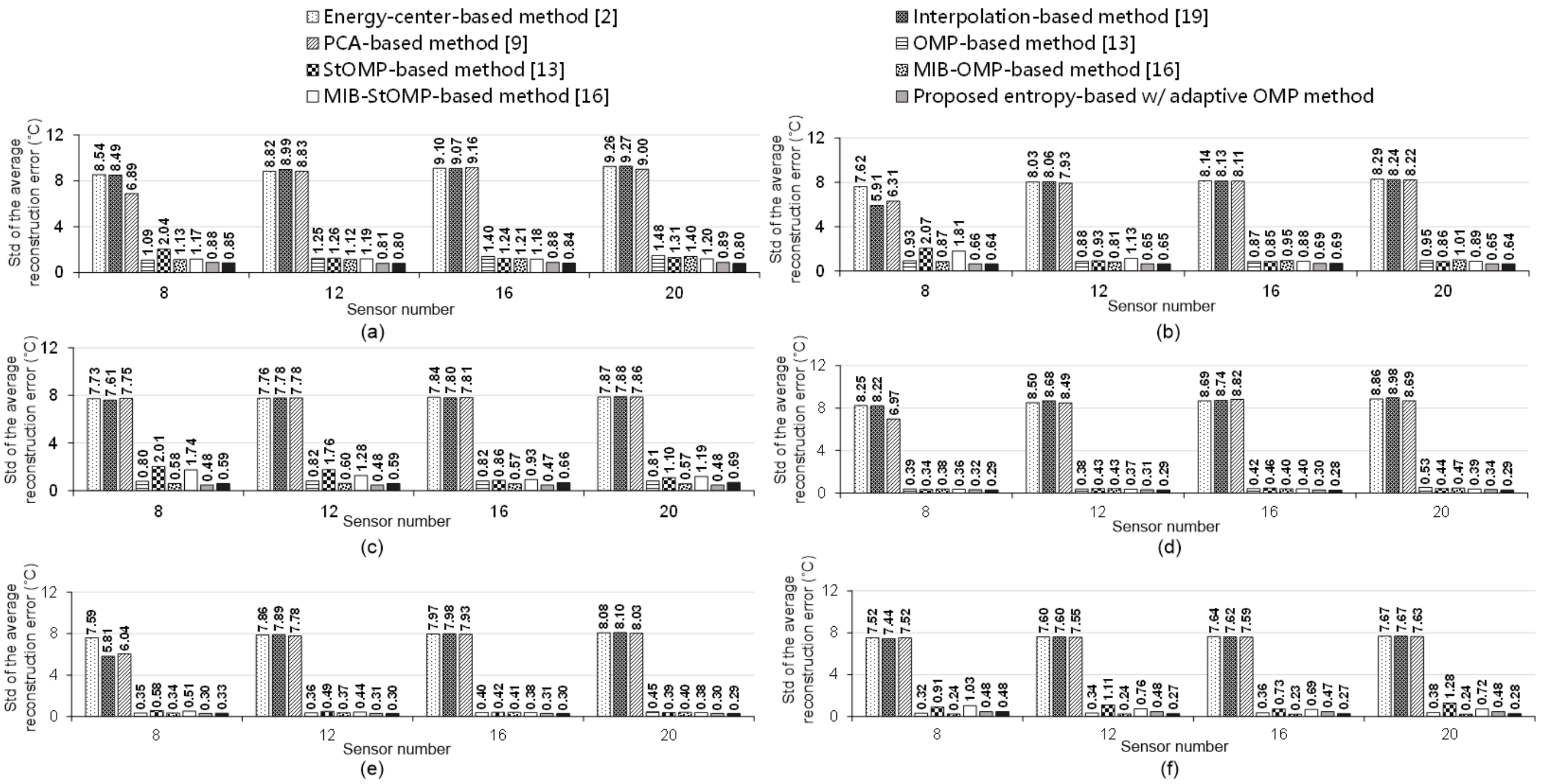

Fig. 8 The standard deviation of average temperature estimation error according to different sensor numbers under an (a) 8×8 grid, (b) 16×16 grid, and (c) 32×32 grid after executing the entire Splash-2 benchmark suite; the standard deviation of average temperature estimation error according to different sensor numbers under an (d) 8×8 grid, (e) 16×16 grid, and (f) 32×32 grid after executing the entire PARSEC benchmark suite.

method [16], and 6) the proposed method.

As mentioned before, some methods, such as methods based on energy center [2], interpolation [19], and PCA [9], require prior knowledge for placing thermal sensors on the multi-core system. Therefore, we adopt the *FFT* application as an initial program within the *SPLASH-2* benchmark suite for offline analysis of related works. Subsequently, we utilize temperature maps from other applications in *SPLASH-2* as online applications. Furthermore, to avoid the extreme results caused by the random placement of the thermal sensors in the MIB-CS-based method, we use the average of 100 simulation results as the evaluation consequence. Fig. 6 displays the comparison of temperature estimation errors using different methods under various numbers of thermal sensors placed within the *SPLASH-2* benchmark suite, and the involved grid sizes are 8-by-8, 16-by-16, and 32-by-32. Fig. 6(a)(c)(e) show the average temperature estimation error, where our proposed entropy-based sensor allocation combined with the adaptive OMP reconstruction method can reduce the full-chip temperature reconstruction error to 18% to 94%, and the adaptive StOMP reconstruction method can reduce it to 19% to 95%. Fig. 6(b)(d)(f) show the maximum temperature reconstruction error, where the adaptive OMP reconstruction method can reduce the error by 22% to 95%, and the adaptive StOMP reconstruction method can reduce it by 20% to 95%. In addition, a similar improvement has been shown in Fig. 7 under the PARSEC benchmark suite. On the other hand, we analyze the standard deviation of the average reconstruction error in Fig. 8. Compared with related works, our proposed methods significantly improve both average and maximum temperature reconstruction errors with smaller standard deviations. The reason is that the proposed entropy-based thermal sensor allocation method aims to cover the temperature behaviors of multi-core systems as much as possible. Besides, the proposed adaptive CS-based reconstruction method further effectively solves the problem of unstable reconstruction quality under time-varying workloads.

To assess the robustness of our method, we introduced Gaussian noise to the sensor readings and examined the stability of our approach compared to related methods. Since all temperature reconstruction techniques involve multiple addition operations (i.e., multiplying the sensor data by certain coefficients), the impact of reconstruction errors resulting from sensing inaccuracies remains consistent. TABLE I shows the analysis of the average temperature reconstruction error with and without temperature sensing noise. Obviously, the temperature reconstruction error is stable, which means that the primary factor influencing the reconstruction error is the coefficients used in the temperature reconstruction method. Because the proposed method adjusts the involved measurement matrix adaptively, we still have a better performance than other related works under noisy sensing data.

Fig. 9 shows the comparison results of the average temperature reconstruction error under different time-varying applications in the *SPLASH-2* benchmark suite according to different numbers of thermal sensor placements. Compared with related works, the methods based on adaptive OMP and adaptive StOMP can achieve lower full-chip temperature reconstruction errors. As mentioned before, the conventional CS-based approach involves a fixed measurement matrix that does not consider the time-varying temperature change. On the other hand, the proposed adaptive CS-based approach can dynamically adjust the involved measurement matrix, thereby reducing the full-chip temperature reconstruction error.

TABLE I
THE AVERAGE TEMPERATURE RECONSTRUCTION ERROR ANALYSIS WITH NOISY TEMPERATURE SENSING DATA UNDER THE SPLASH-2 BENCHMARK SUITE

| Grid Size = 8 x 8 | | | | | |
|---|---|---|---|---|---|
| **Method** | **Noise** | **Sensor number** | | | |
| | | **8** | **12** | **16** | **20** |
| **Energy-center-based method [2]** | **w/o** | 12.13 | 11.21 | 10.73 | 10.21 |
| | **w** | 12.14 | 11.21 | 10.73 | 10.21 |
| **Interpolation-based method [19]** | **w/o** | 11.81 | 11.90 | 10.97 | 9.89 |
| | **w** | 11.83 | 11.91 | 10.98 | 9.89 |
| **PCA-based method [9]** | **w/o** | 9.34 | 11.21 | 10.64 | 9.26 |
| | **w** | 9.35 | 11.22 | 10.65 | 9.27 |
| **OMP-based method [27]** | **w/o** | 1.79 | 1.64 | 1.00 | 0.87 |
| | **w** | 2.17 | 1.67 | 1.04 | 1.03 |
| **StOMP-based method [13]** | **w/o** | 1.37 | 1.10 | 0.92 | 0.78 |
| | **w** | 1.50 | 1.26 | 1.05 | 0.90 |
| **MIB-OMP-based method [16]** | **w/o** | 1.73 | 1.65 | 1.00 | 0.87 |
| | **w** | 1.88 | 1.89 | 1.50 | 0.97 |
| **MIB-StOMP-based method [16]** | **w/o** | 1.36 | 1.15 | 0.91 | 0.77 |
| | **w** | 1.51 | 1.33 | 0.92 | 0.86 |
| **Proposed entropy-based w/ adaptive OMP method** | **w/o** | 0.84 | 0.68 | 0.73 | 0.63 |
| | **w** | 0.91 | 0.77 | 0.80 | 0.74 |
| **Proposed entropy-based w/ adaptive StOMP method** | **w/o** | 0.66 | 0.57 | 0.66 | 0.62 |
| | **w** | 0.72 | 0.65 | 0.74 | 0.69 |
| **Grid Size = 16 x 16** | | | | | |
| **Method** | **Noise** | **Sensor number** | | | |
| | | **8** | **12** | **16** | **20** |
| **Energy-center-based method [2]** | **w/o** | 12.73 | 13.73 | 13.02 | 13.39 |
| | **w** | 12.74 | 13.73 | 13.03 | 13.39 |
| **Interpolation-based method [19]** | **w/o** | 10.17 | 13.71 | 13.00 | 12.70 |
| | **w** | 10.18 | 13.72 | 13.02 | 12.72 |
| **PCA-based method [9]** | **w/o** | 10.33 | 14.18 | 13.73 | 13.39 |
| | **w** | 10.33 | 14.19 | 13.73 | 13.39 |
| **OMP-based method [27]** | **w/o** | 1.22 | 0.97 | 0.94 | 1.03 |
| | **w** | 1.02 | 1.07 | 1.30 | 1.21 |
| **StOMP-based method [13]** | **w/o** | 3.36 | 1.11 | 0.96 | 0.94 |
| | **w** | 1.01 | 1.37 | 1.10 | 1.12 |
| **MIB-OMP-based method [16]** | **w/o** | 1.04 | 0.93 | 1.08 | 1.16 |
| | **w** | 1.06 | 1.11 | 1.20 | 1.14 |
| **MIB-StOMP-based method [16]** | **w/o** | 4.86 | 1.02 | 0.94 | 1.01 |
| | **w** | 1.13 | 1.12 | 1.15 | 1.13 |
| **Proposed entropy-based w/ adaptive OMP method** | **w/o** | 0.81 | 0.76 | 0.74 | 0.64 |
| | **w** | 0.88 | 0.84 | 0.83 | 0.78 |
| **Proposed entropy-based w/ adaptive StOMP method** | **w/o** | 0.79 | 0.75 | 0.79 | 0.62 |
| | **w** | 0.85 | 0.82 | 0.81 | 0.75 |
| **Grid Size = 32 x 32** | | | | | |
| **Method** | **Noise** | **Sensor number** | | | |
| | | **8** | **12** | **16** | **20** |
| **Energy-center-based method [2]** | **w/o** | 13.52 | 13.46 | 13.64 | 13.58 |
| | **w** | 13.53 | 13.47 | 13.65 | 13.60 |
| **Interpolation-based method [19]** | **w/o** | 13.15 | 14.24 | 14.21 | 13.60 |
| | **w** | 13.16 | 14.25 | 14.22 | 13.61 |
| **PCA-based method [9]** | **w/o** | 13.91 | 13.82 | 13.73 | 13.59 |
| | **w** | 13.92 | 13.84 | 13.74 | 13.60 |
| **OMP-based method [27]** | **w/o** | 0.95 | 1.02 | 1.05 | 0.99 |
| | **w** | 1.01 | 1.08 | 1.07 | 1.11 |
| **StOMP-based method [13]** | **w/o** | 6.12 | 1.29 | 0.93 | 1.15 |
| | **w** | 2.87 | 1.30 | 1.07 | 1.31 |
| **MIB-OMP-based method [16]** | **w/o** | 0.65 | 0.67 | 0.64 | 0.62 |
| | **w** | 0.67 | 0.65 | 0.65 | 0.64 |
| **MIB-StOMP-based method [16]** | **w/o** | 3.13 | 3.04 | 0.88 | 1.10 |
| | **w** | 1.75 | 1.25 | 1.17 | 1.19 |
| **Proposed entropy-based w/ adaptive OMP method** | **w/o** | 0.61 | 0.60 | 0.58 | 0.57 |
| | **w** | 0.63 | 0.61 | 0.59 | 0.57 |
| **Proposed entropy-based w/ adaptive StOMP method** | **w/o** | 0.79 | 0.82 | 0.77 | 0.82 |
| | **w** | 0.84 | 0.88 | 0.85 | 0.87 |

While the consequences of thermal sensor placement using adaptive CS-based reconstruction methods bear resemblance, the processing time for the StOMP-based method [13] tends to escalate as the size of the grid increases. TABLE II shows the comparison of the execution time under different gird sizes, and the number of allocated thermal sensors is eight. In comparison to the traditional StOMP method [13], the proposed adaptive OMP method facilitates a processing time reduction by 10% to 18%. Moreover, our proposed adaptive StOMP method achieves a notable decrease in processing time, between 19% and 27%. Given that the thermal profile of a chip undergoes changes over hundreds of milliseconds [24], the proposed methodology guarantees efficient monitoring of alterations in temperature distribution. However, due to the proposed method requiring dynamic adjustment of the measurement matrix at each iteration, its performance in processing time is not as efficient as the MIB-based method [16]. This slight increase in time cost helps to reduce the reconstruction errors.

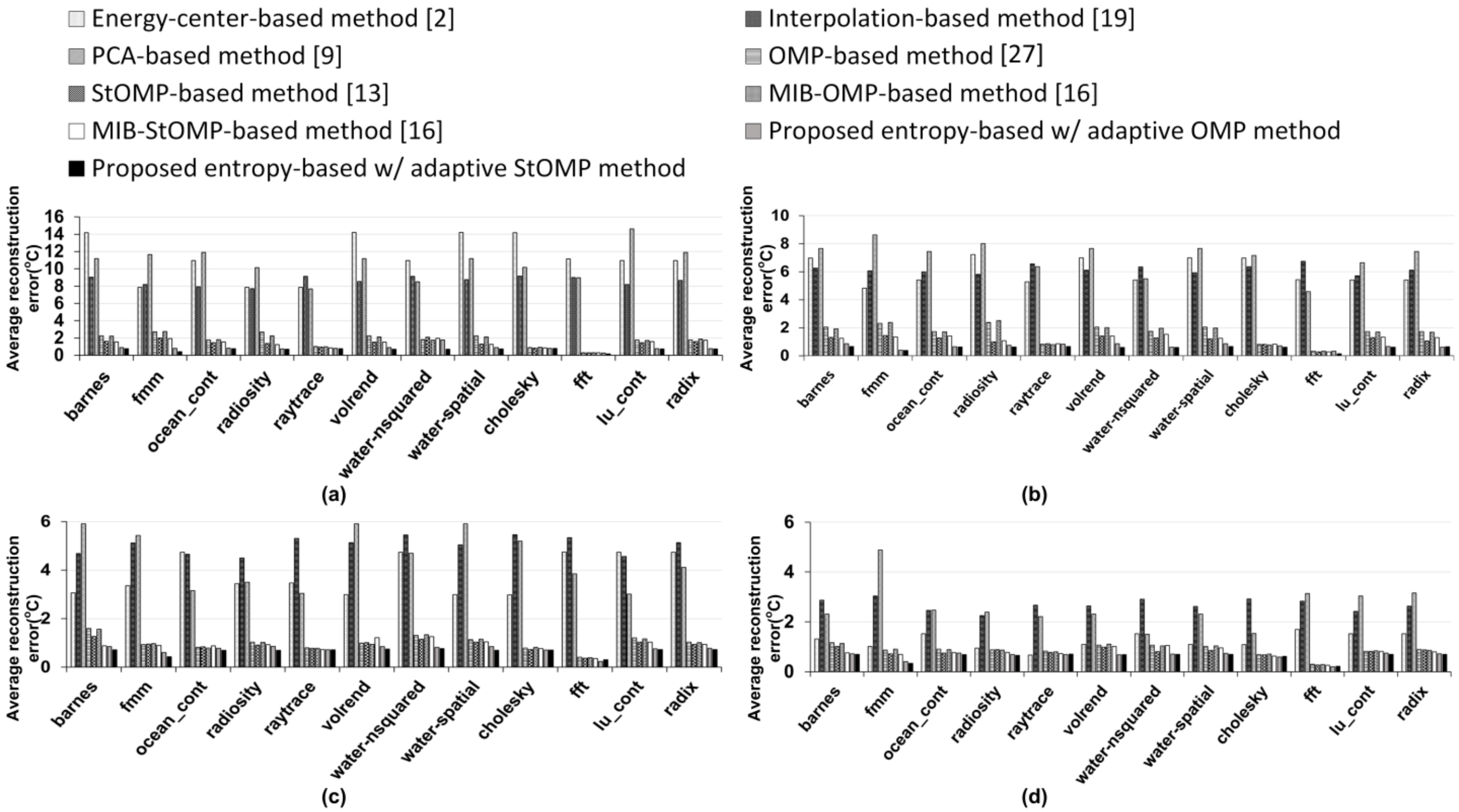

Fig. 9 The comparison of average temperature reconstruction across time-varying applications in SPLASH-2 using different methods, featuring (a) 8 sensors, (b) 12 sensors, (c) 16 sensors, and (d) 20 sensors deployed.

TABLE II
THE COMPUTATIONAL COMPLEXITY AND EXECUTION TIME.

| | Computational Complexity | Execution Time under Different Grid Size | | |
|---|---|---|---|---|
| | | *8×8* | *16×16* | *32×32* |
| **StOMP[13]** | $O(N^2)$ | 0.031 sec. | 0.041 sec. | 0.051 sec. |
| **MIB-based OMP[16]** | O(N) | 0.024 sec. (-22%) | 0.030 sec. (-26%) | 0.037 sec. (-27%) |
| **MIB-based StOMP[16]** | O(N) | 0.021 sec. (-32%) | 0.025 sec. (-39%) | 0.030 sec. (-41%) |
| **Adaptive OMP** | O(N) | 0.028 sec. (-10%) | 0.036 sec. (-12%) | 0.042 sec. (-18%) |
| **Adaptive StOMP** | O(N) | 0.025 sec. (-19%) | 0.031 sec. (-24%) | 0.037 sec. (-27%) |

### *B. Verification on FPGA*

To analyze the feasibility of the proposed method on a practical multi-core system platform, we use a FLIR Lepton infrared camera to collect thermal maps of the Xilinx ZC706 FPGA at runtime for verification. The mobile phone can collect a series of thermal maps on the multi-core system through

TABLE III
THE SPECIFICATION OF THE INVOLVED FLIR LEPTON INFRARED CAMERA

| | |
|---|---|
| **Spectral range** | Longwave infrared, 8μm to 14μm |
| **Array format (Resolution)** | 80 x 60 |
| **Frame rate** | 8.6 Hz |
| **Thermal sensitivity** | < 50 mK (0.050°C) |
| **Scene dynamic range** | High Gain Mode: -10°C to 140°C<br>Low Gain Mode: -10°C to 450°C |
| **Radiometric accuracy** | High gain: Greater of ±5°C or 5%<br>Low gain: Greater of ±10°C or 10% |

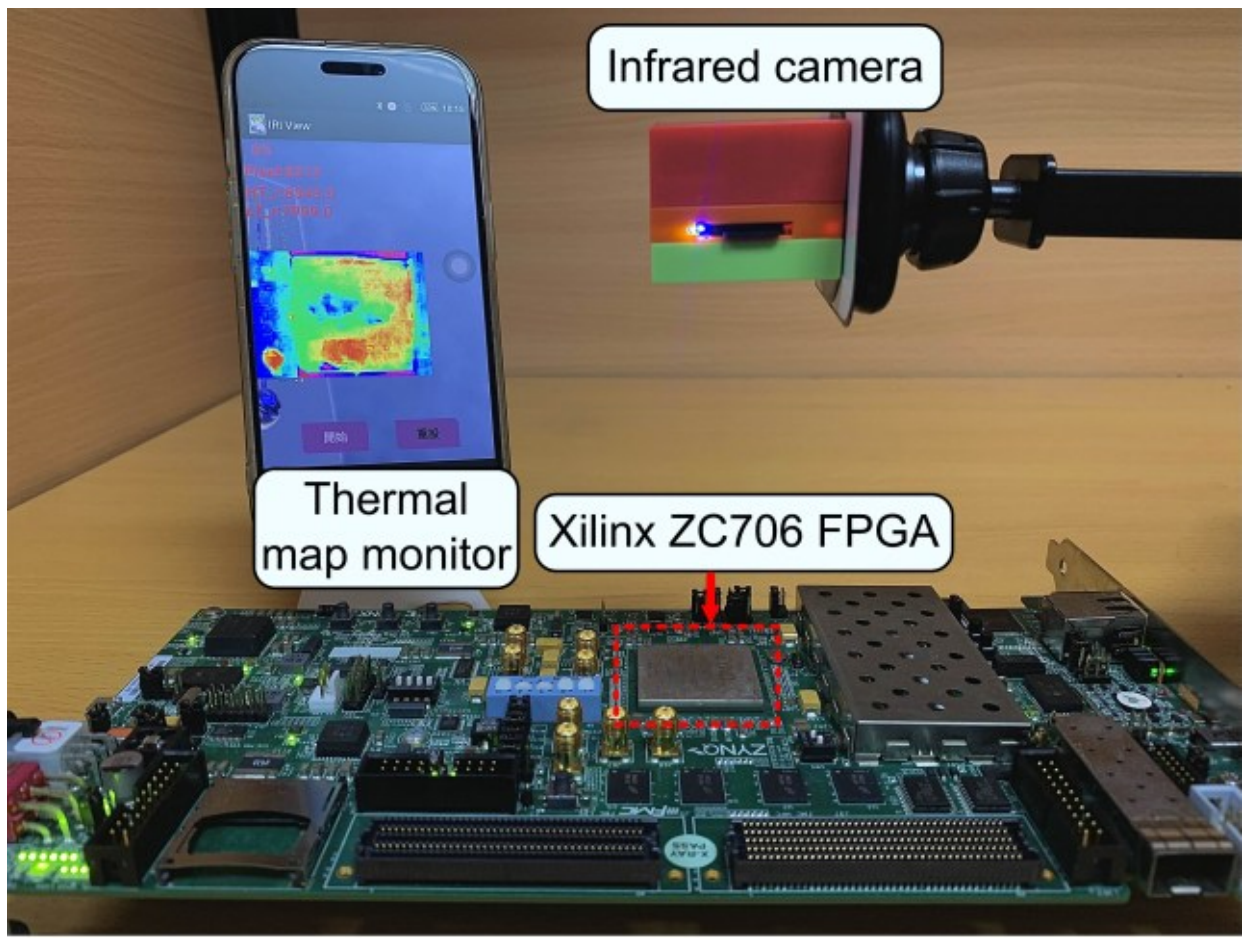

Fig. 10 The image of the practical verification platform.

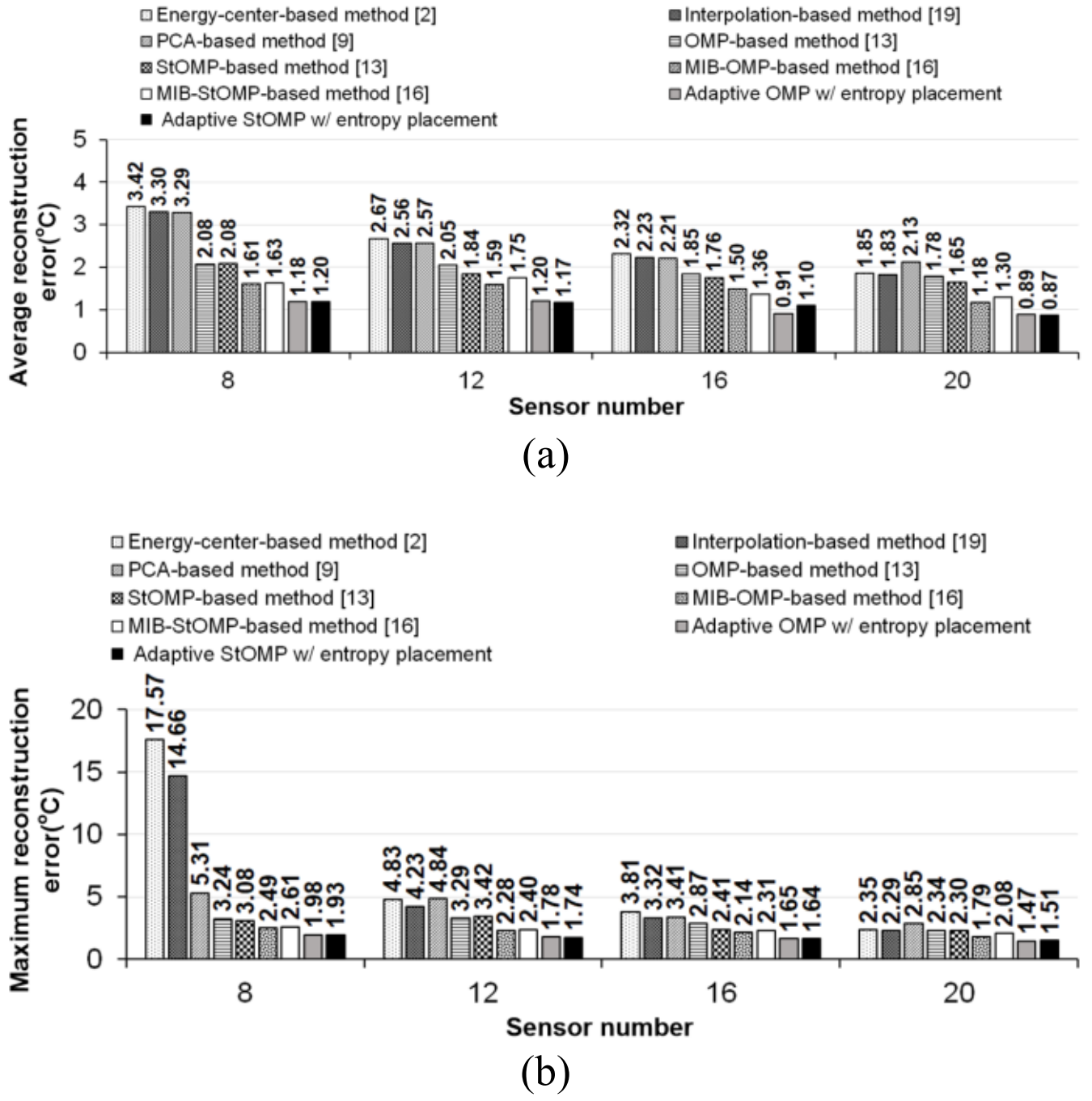

Fig. 11 The (a) average reconstruction error; (b)maximum reconstruction error comparison under a practical multi-core system-based LeNet CNN design.

Bluetooth connected to the FLIR Lepton infrared camera, and the actual verification system platform is shown in Fig. 10.The detailed specifications of the thermal infrared camera used in this experiment are shown in TABLE III. Given that the thermal profile on chips requires hundreds of milliseconds to change [24], measurements are updated every 116 milliseconds to instantly respond to changes in system temperature [26].

In this experiment, we execute a LeNet CNN on the FPGA and compute the MNIST dataset. In this way, we can use real temperature data on a multicore system to evaluate the proposed method and the related works, as shown in Fig. 11. The experimental results can be categorized into three sections for assessment. Firstly, compared with non-CS methods [2][9][19], the proposed method can reduce the average error by 55% to 88% and the maximum error by 34% to 89%. Second, compared with the traditional CS method [13], the proposed method can reduce the average error by 35% to 51% and the maximum error by 31% to 49%. Finally, compared with the MIB-CS-based methods [16], the proposed method can reduce the average error by 19% to 39% and the maximum error by 16% to 29%. The reason is that the proposed method can place the thermal sensors to cover as many different temperature behaviors as possible. On the other hand, the involved measurement matrix can be updated based on the real-time temperature reconstruction error to fit the current temperature behavior at runtime.

### *C. Implementation result and hardware efficiency*

To analyze the area overhead associated with the traditional CS reconstruction method, the MIB-CS reconstruction method [16], and the proposed adaptive CS temperature reconstruction architecture, we utilized the TSMC 40nm technology process to implement each of them. For a fair comparison, we utilized the design parameters in [28] and assumed that the sensing range of the thermal sensor is from -10°C to 110°C. Additionally, the sensing resolution was set to 0.32°C. Our implementation involved 16-bit fixed-point operations with 8 bits allocated to the integer part and another 8 bits to the fractional part, ensuring that the method meets the temperature range requirements of multi-core systems. Besides, the grid size is 8-by-8, and the number of thermal sensors is eight.

TABLE IV shows the comparison of the hardware cost between the proposed method and other related works. Because our proposed approach is based on the MIB-based method and uses LMS-based adaptive filter theory to adjust the involved measurement matrix, the area and power of the proposed method are larger than those of the MIB-based methods [16]. However, compared with the conventional CS-based methods [13][27], the area overhead is small due to the involved MIB-based matrix operation. To consider the design trade-off between area, latency, power, and full-chip temperature reconstruction error, we design a cost function for hardware efficiency (HE), formulated to

$$HE = \frac{1}{\text{area} \times \text{latency} \times \text{power} \times \text{error}}. \tag{7}$$

TABLE IV
THE SYNTHESIS RESULT OF THE DIFFERENT CS-BASED TEMPERATURE RECONSTRUCTION METHODS

| | OMP[27] | StOMP[13] | MIB-based OMP[16] | MIB-based StOMP[16] | Proposed Adaptive OMP | Proposed Adaptive StOMP |
|---|---|---|---|---|---|---|
| **Area (μm²)** | | | | | | |
| **Temperature reconstruction computing unit** | 169,997 | 172,025 | 96,565 | 109,120 | 90,953 | 112,000 |
| **LMS-based matrix updater** | - | - | - | - | 22,646 | 12,382 |
| **Memory** | 29,995 | 29,714 | 48,941 | 38,468 | 55,535 | 46,849 |
| **Total area** | 199,992 | 201,739 | 145,506 | 147,588 | 169,134 | 171,230 |
| **Power (mW)** | | | | | | |
| **Temperature reconstruction computing unit** | 2.39 | 2.42 | 1.92 | 2.11 | 1.50 | 1.71 |
| **LMS-based matrix updater** | - | - | - | - | 0.36 | 0.32 |
| **Memory** | 0.27 | 0.27 | 0.42 | 0.29 | 0.53 | 0.42 |
| **Total Power (mW)** | 2.66 | 2.69 | 2.34 | 2.40 | 2.40 | 2.45 |
| **Latency (ns)** | 264,692 | 209,557 | 198,741 | 148,221 | 198,781 | 148,261 |
| **Average Temperature Reconstruction Error (°C)** | 1.79 | 1.37 | 1.73 | 1.36 | 0.84 | 0.66 |
| **Normalized HE index** | 1 | 1.62 | 2.15 | 3.53 | 3.72 | 6.14 |

We set the HE index of the OMP method as the baseline result. Compared with other related works, the proposed adaptive CS-based method by using OMP algorithm can improve the HE by 5% to 272%. Besides, the proposed adaptive CS-based method by using StOMP can improve the HE by 74% to 514% over other related works. This demonstrates that the adaptive CS-based method can achieve significantly more accurate reconstruction with only a slight increase in hardware overhead.

## V. Conclusion

This paper addresses the challenges of thermal sensor allocation and full-chip temperature reconstruction in multi-core systems by leveraging an entropy-based sensor placement strategy and an adaptive compressive sensing approach. By selecting sensor locations that capture diverse thermal behaviors and dynamically adjusting the measurement matrix, our method significantly enhances the accuracy of the full-chip temperature reconstruction. Experimental results demonstrate that our approach reduces full-chip temperature reconstruction error by 18% to 95%. In addition to the full-chip temperature reconstruction efficiency enhancement, our proposed method improves hardware efficiency by 5% to 514% over the related works. These findings highlight the potential of our method for more effective dynamic temperature management in future high-performance multi-core systems.

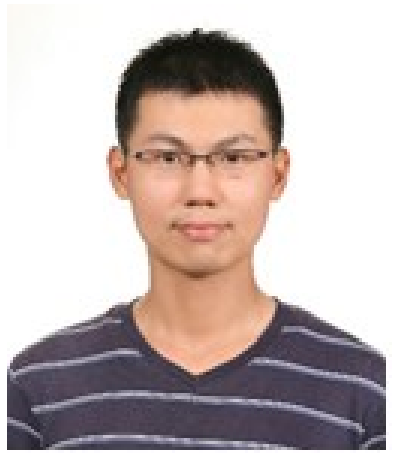

**Kun-Chih (Jimmy) Chen (IEEE S'10-M'14-SM'21)** is currently an Associate Professor and Electric Junior Chair Professor at the Institute of Electronics, National Yang Ming Chiao Tung University (NYCU). His research interests include Multiprocessor SoC (MPSoC) design, Neural network learning algorithm design, Reliable system design, VLSI/CAD design, and Smart manufacturing. Prof. Chen heads many services in IEEE Circuits and Systems Society, such as IEEE JETCAS Guest Editor, General Chair of NoCArc 2020. Prof. Chen has received several prestigious national and international awards, including the Ta-You Wu Memorial Award of NSTC (i.e., Early Career Award in Taiwan), IEEE CASS Continuing Education Featuring Selected Conference Tutorial, IEEE Tainan Section Best Young Professional Member Award, Taiwan IC Design Society Outstanding Young Scholar Award, etc. Under his leadership, his research team receives the first IEEE ISCAS Best Student Paper Award and IEEE TVLSI Best Paper Award in Taiwan. He is an IEEE senior member.

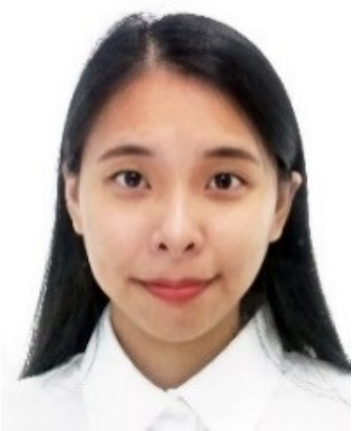

**Chia-Hsin Chen** received her B.S. degree from National Chi Nan University, Taiwan, in Computer Science and Information Engineering in 2020 and the M.S. degree from the Department of Computer Science and Engineering, National Sun Yat-sen University, Kaohsiung, Taiwan, in 2022.

She is currently an Engineer with Phison Electronics. Her research interests focus on thermal-aware multicore system design.

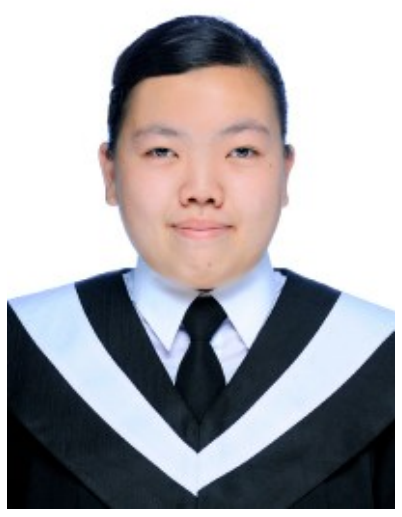

**Lei-Qi Wang** received her B.S. degree in Computer and Communication from National Pingtung University, Taiwan, in 2021. She is currently pursuing her M.S. degree in the Department of Computer Science and Engineering at National Sun Yat-sen University, Taiwan. Her research interests focus on thermal sensor placement for multi-core systems.

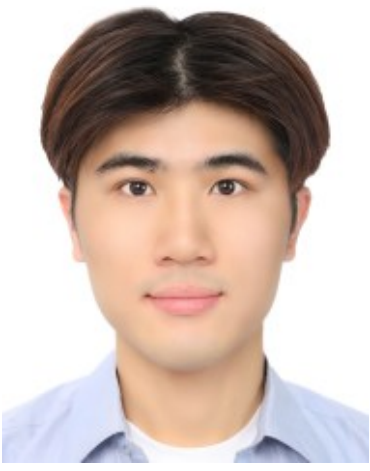

**Chun-Chieh Wang** received his B.S. degree in Electrical Engineering from National Cheng Kung University, Taiwan, in 2023. He is currently pursuing his M.S. degree at the Institute of Electronics, National Yang Ming Chiao Tung University, Taiwan. His research interests focus on thermal-aware multi-core systems.